\documentclass[12pt, english]{article}

\usepackage{amsmath,amssymb,color,float}
\usepackage{bm}
\usepackage{stmaryrd}
\usepackage{graphicx,psfrag,epsf}
\usepackage[authoryear]{natbib}
\usepackage{fullpage,setspace}
\usepackage{subcaption}
\usepackage{changepage}
\usepackage{listings}
\usepackage{colortbl}
\usepackage{xcolor}
\usepackage{ragged2e}
\usepackage{tabularx}
\usepackage{authblk}
\usepackage{booktabs}
\usepackage{placeins}
\usepackage{url}

\usepackage{lineno}

\newcommand{\Prob}{\mathbb{P}}

\title{Improving ecological inference and uncertainty quantification from camera trap data through the fusion of AI confidences and manual annotations}

\author[1]{Adira~Cohen}
\author[1]{Erin~M.~Schliep}
\author[2,3]{Mohammad~Alyetama}
\author[2]{Matthew~Snider}
\author[2,4]{Roland~Kays}
\affil[1]{North Carolina State University, Raleigh, NC, USA}
\affil[2]{Department of Forestry and Environmental Resources, North Carolina State University, Raleigh, NC, USA}
\affil[3]{Department of Psychology, Neuroscience and Behavior, University of Nebraska Omaha, Omaha, NB, USA}
\affil[4]{North Carolina Museum of Natural Sciences, Raleigh, NC, USA}

\begin{document}

\maketitle

\begin{abstract}
Camera traps have become an important tool in ecological research, enabling large-scale, noninvasive monitoring of wildlife populations and behavior. 
By automatically recording animals as they pass within view, these devices generate massive image datasets with minimal field effort. 
This data richness introduces a new bottleneck when translating the images into usable information due to time and effort required for human annotation. Artificial intelligence (AI) has recently been integrated into the workflow to improve efficiency.
However, the data procured from AI approaches are of a different nature, necessitating new statistical methods. 
We develop a new Bayesian hierarchical data-fusion model that combines the strengths of human annotations and AI predictions.
The benefits of our approach are an ability to provide uncertainty quantification as well as improved inference and predictive power, which we demonstrate through simulation.
We apply our model to an AI analysis of the body condition of white-tailed deer (\emph{Odocoileus virginianus}) from camera trap images from North Carolina to study the relationship between health and their environment.
Our analysis derived novel ecological inference compared to a more traditional approach using the same data. 
We find that bucks in rut have higher (healthier) body condition than other deer and that green, open habitats are correlated with high body condition.
\end{abstract}

\section{Introduction}

Camera traps and automated sound recording units have revolutionized the sampling of wildlife populations due to their low cost and noninvasive nature.
The recordings of species presence can be used for distribution studies and population monitoring via occupancy and abundance modeling \cite[e.g., ][]{Goldstein2024, Schliep2024}.
Photo captures can also be used to study animal body condition and diseases \cite[e.g., ][]{Muneza2019, Ringwaldt2025}.
Improved statistical approaches enable the use of data from these devices to conduct inference at larger scales.
The increased scale is beneficial for ecology and conservation efforts but has led to challenges in processing the larger volumes of data \cite[][]{Leorna2022, Delisle2021, Terry2020}.
Specifically, identifying the species --- or otherwise characterizing the data from the image or sound --- requires extensive manual work.

Artificial intelligence (AI) is useful for quickly processing these large datasets, enabling the continued increase of the scale of surveys.
The use of convolutional neural networks, a popular type of AI model, for image classification has grown rapidly over the past decade \cite[][]{Oliveira2025}. AI models using camera trap data have been shown to be effective for filtering blanks \cite[e.g.,][]{Mulero-Pazmany2025, Henrich2024} and species identification 
\cite[e.g.,][]{Lonsinger2024, Mulero-Pazmany2025},
as well as disease detection \cite[e.g.,][]{Ringwaldt2025} and distance estimation \cite[e.g.,][]{Henrich2024}. 

While AI predictions provide a solution to the annotation bottleneck, statistical methods using AI-integrated camera trap datasets have not kept pace.  
This is in part due to a lack of standardization in the output from AI models, including how AI confidences are used and interpreted \cite[][]{Cowans2024}. 
Statistical models that do exist \cite[e.g.,][]{Rhinehart2022, Cole2022} tend to only be applicable to unrealistically simplified study scenarios.
Many approaches \citep[e.g.,][]{Cowans2024, Ringwaldt2025} employ thresholding, which removes observations with low-confidence scores.
This not only results in a loss of information -- potentially entire survey sites -- but could introduce bias if low confidences are correlated with the response variable of interest. 

Properly integrating AI predictions into statistical models is a challenging task. 
To this end, a variety of approaches have been proposed with differing advantages and disadvantages.
First, the typical approach is to only consider AI predictions \citep[e.g.,][]{Schwob2025} and discard low confidence results \cite[e.g.,][]{Lin2021},
mirroring models for compositional data \cite[e.g.,][]{Feng2017}. 
The disadvantage of this approach is that some data are discarded, and that AI accuracy tends to be lower than human accuracy for all but simple tasks \cite[][]{Kitzes2025}. 
Moreover, unlike statistical models, AI confidences are not generally \emph{calibrated} (i.e., converted to probabilities). Building probabilistic models that properly quantify uncertainty using AI confidences necessitates the use of additional data, such as manual annotations \cite[][]{Dussert2025, Kitzes2025}.

Other analyses fit models separately to 
AI predictions and manual annotations \cite[e.g.,][]{Ringwaldt2025, Henrich2024}, which allows for some uncertainty quantification through model comparison. These studies show that although AI predictions have more noise than manual annotations, they can produce comparable results given enough data. 
While an improvement, this approach is not ideal because it does not collectively leverage the unique strengths of each dataset. 
There is no direct sharing of information between models, so that the models fit on the AI predictions lack uncertainty quantification and the models fit on the manual annotations are limited by study scale.

A third approach uses manual annotations to calibrate the AI confidences, which converts the AI confidences into probabilities \cite[e.g.,][]{Ware2023, Wood2024}. 
This provides a valuable improvement by using the strengths of each data type to quantify uncertainty in the model. 
However, proper uncertainty propagation is challenging in this two-stage procedure. 

Lastly, models that jointly use both AI predictions and manual annotations, often referred to as data fusion models, have recently been developed  \cite[e.g.,][]{Rhinehart2022, Doser2021}. 
This approach is common in other applications when there is one abundant but poor-quality data source and a second source that is scarce but of high quality \cite[e.g.,][]{Zaiats2024}.
In this case the data sources are also different in nature, as manual annotations are ordinal (integers) while the AI outputs a compositional vector conveying its confidence of each possible ordinal score.
Joint modeling allows the strengths and weaknesses of each data source to complement each other, where the manual annotations provide calibration and uncertainty quantification, and the AI confidences provide massive amounts of inexpensive data.
However, the disparate data types necessitate new statistical methods to obtain inference, make predictions, and quantify uncertainty. 

Recent efforts that explore the fusion of AI classification data with manual annotations include \cite{Rhinehart2022} for occupancy modeling and \cite{Spence2025} for abundance modeling. In occupancy modeling, manual annotations provide binary observations of presence/absence data, while AI confidences are continuous measures of presence.  \cite{Rhinehart2022} model each of these data sources conditionally given a true latent variable for presence/absence. Their model is defined at the site level and explores variation in occupancy as a function of environmental drivers. \cite{Spence2025} have two observations of abundance (count data) also at the site level that result from aggregating manual annotations and AI classifications of images. Importantly, the classifications are obtained by extracting categorical predictions from the continuous confidences produced from an AI model. Models for presence/absence and abundance data are more customary in the literature. Our data being ordinal and our interest in deer body condition at the image level require the development of new inferential and predictive models.

The current approaches are limited in that they either discard valuable sources of uncertainty --- such as manual annotations, the distribution of AI confidences, or via thresholding --- or do not fully leverage the available data to improve inference and prediction.
To address these concerns, we propose a Bayesian hierarchical data-fusion model which combines the strengths of human annotations and AI predictions.
The benefits of our approach are 
an ability to provide uncertainty quantification and improved inference and prediction power.
We incorporate the nature of the data, in particular modeling the AI outputs as confidences rather than probabilities.
Our analysis uses camera trap data from the Candid Critters project \cite[][]{Lasky2021}, which encompasses all of North Carolina. 
We are interested in using the images from this study to understand the relationship between the body condition of white-tailed deer and their environment.

In Section~\ref{sec:DataDescription} we describe the Candid Critters dataset and the two sources of data -- human annotations and AI confidences. We present our proposed data fusion model in Section~\ref{sec:Model}, and include details on model fitting and inference. In Section~\ref{sec:Simulation} we include a simulation study to demonstrate the strength of our approach. In Section~\ref{sec:CaseStudy}, we present the results from fitting our model to the Candid Critters dataset. We conclude with a discussion of our model and results in Section~\ref{sec:Discussion}.

\section{Description of the data} \label{sec:DataDescription}

\subsection{Camera traps and Candid Critters}

\begin{figure}[ht]
    \centering
    \includegraphics[width=1\linewidth]{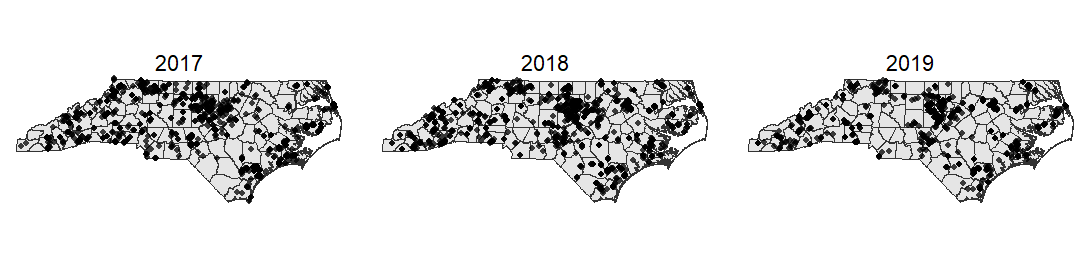}
    \caption{Deployment locations for the Candid Critters dataset spanning North Carolina after data cleaning separated by year. There are at total of $104464$ images representing $17471$ sequences across $2766$ deployments.}
    \label{fig:deployments}
\end{figure}

Candid Critters was an organized citizen science camera trapping project across North Carolina with participants from all $100$ counties \cite[][]{Lasky2021}. 
There are $4295$ deployments (camera placements) from $2016$ to $2019$, the locations of which are shown in Figure~\ref{fig:deployments}. 
The cameras were placed on both private and public land with retrievals collected from all seasons. 
The cameras were set at least $200 \text{m}$ apart without bait, and were pointed away from high-traffic regions such as a bird-feeders, game-trails, or roads. 
The cameras used were motion-sensitive infrared flash trail cameras, meaning that they are triggered by movement of heat (such as a mammal walking in front of the camera) and are able to take images during the day or night. 
When triggered, the cameras take $5$ photographs at approximately $1$ second intervals for each trigger, and retrigger immediately if the animal is still present, resulting in a \emph{sequence} of images. 
Over the span of the Candid Critters project, $2.2$ million images of wildlife were collected, $56\%$ of which were white-tailed deer (\emph{Odocoileus virginianus}, hereafter ``deer''), our species of interest. 
The large number of images of deer creates problems of human effort when annotating images, which motivates our analysis. 

Within each sequence there are multiple images, each of which may either be blank or contain one or more individuals. 
Each individual animal is given a bounding box to identify its location in the image.
For our analysis, species and group size were already manually annotated for each sequence.
With group sizes of $1$, the sequence is a repeated survey of a single individual.
For larger groups, it is impossible to accurately track individuals across frames, so we do not include these in our analysis.

\subsection{Deer body condition score and manual annotations}

\begin{figure}[ht]
    \centering
    \includegraphics[width=\linewidth]{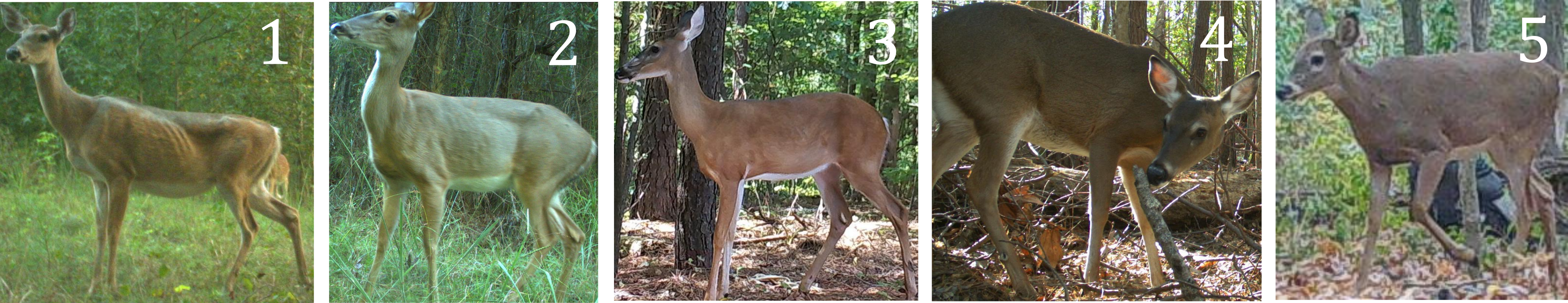}
    \caption{Images of white-tailed deer taken by camera traps labeled with their body condition score.}
    \label{fig:ex_health}
\end{figure}

Body Condition Score (BCS) is a measure designed to assess certain health indicators of white-tailed deer, such as the appearance of the ribs, spine, and belly \cite[][]{Smiley2017, McGraw2022}. 
The BCS takes on ordinal integer values of $1$, $2$, $3$, $4$, or $5$, where $1$ is the least healthy and $5$ is the most healthy. 
A deer with a BCS of $1$ appears extremely thin and bony whereas a deer with a BCS of $5$ appears plump. 
See Figure~\ref{fig:ex_health} for an example of each ordinal category. 

The human annotated data used in our analysis were obtained from a team of researchers who were trained in scoring body condition.
The annotators scored unique partitions of images meaning each image was scored by at most one annotator, and images were viewed independently without considering other images in the sequence. 
This is the slowest yet most reliable method for scoring body condition, although the annotators are fallible and may have some individual classification bias. 

\subsection{AI model}

An image can be assigned a set of confidences by an artificial intelligence (AI) model. 
This method is much faster than manual annotations but less reliable. 
In addition to inherent classification error through AI, the precision of the AI confidences may vary depending on factors such as light exposure. 
The data used in our analysis came from applying an AI model trained to score the body condition of deer \cite[][]{Alyetama} to images of deer from the Candid Critters dataset.
The first AI sub-model accepts individual images of deer as input and filters them using a convolutional neural network to predict image quality.
High-quality images are passed to another AI sub-model which classifies the BCS of each deer.
The BCS sub-model treats BCS as a categorical random variable (i.e., unordered) and uses standard softmax cross-entropy loss. 
For each image the AI model provides compositional data consisting of a vector of confidences of the BCS being in each ordinal category. 
The confidences must sum to $1$ and higher values denote higher confidence. 
For example, an AI confidence of $[ 0, 0.1, 0.7, 0.2, 0 ]$ implies that the model is most confident that the deer has BCS $3$, with lower confidence for scores of 2 and 4.

The AI model significantly increases the amount of data while decreasing the time frame of the project with respect to both human effort and amount of time until the data are available.
After training on $2869$ manually annotated images, the AI model scored all $104464$ images of solitary deer in the Candid Critters dataset in approximately $40$ minutes with almost no human effort, whereas it would have taken over 5 months of dedicated human effort to score these images by hand. 
These images represent $17471$ sequences at $2766$ deployments, which contain $92$ manual annotations and $104464$ AI confidences.
Importantly, no images used in our analysis were used in the training of the AI model. 

\subsection{Ecology model}

\begin{figure}[ht]
    \centering
    \includegraphics[width=\linewidth]{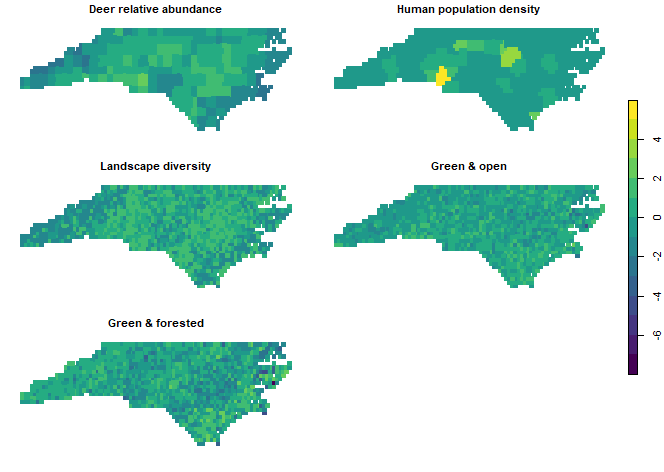}
    \caption{Maps of environmental covariates after centering and scaling, where the NDVI values used to calculate ``green \& open'' and ``green \& forested'' are depicted for November 2018 as an example.}
    \label{fig:covariate_maps}
\end{figure}

We are interested in understanding how the body condition of deer is related to certain individual and environmental drivers.
Life stages of deer are known to be important drivers of variation in body structure. 
We classify deer as being a buck during rut if they are a male deer captured in October to December. 
If the deer is captured during this time period then the deer is sexed using an AI model \cite[][]{Alyetama}, and being a buck during rut or not is equivalent to being male versus female.
Otherwise, all deer are classified as not being a buck during rut, since deer without antlers are difficult to sex.
Bucks during rut are expected to have the highest body condition because they spend the previous months gaining weight to compete for does.
We also consider other landscape variables such as deer relative abundance at 10 km resolution \cite[from][]{Kays2024}, county-level human population density \cite[from][]{CountyPop}, landscape diversity at 1km resolution \cite[from][]{Jung2020}, normalized difference vegetation index (NDVI) at 1km resolution \cite[from][]{MoveBank, Didan2021}, and percent of tree canopy cover at 30m resolution \cite[from][]{Sexton2013}. 
All environmental covariates are constant in time except NDVI, which is sampled in monthly intervals.
To differentiate between high NDVI caused by forest (which may not be accessible by deer) from short vegetation, NDVI and tree canopy are combined using principal component analysis (PCA). 
The first component measures how green (high NDVI) and open (low tree canopy value) the habitat is and the second component measures how green and forested (high tree canopy value) the habitat is. 
We expect that deer will have higher body condition in areas with green and open habitat because higher values indicate larger amounts of accessible vegetation. 
The spatial distributions of the environmental drivers across North Carolina are given in Figure~\ref{fig:covariate_maps}.
The environmental drivers are centered and scaled before fitting the model.

\section{Fusion model specification and inference} \label{sec:Model}

\subsection{Model framework}
\label{sec:Model.Framework}

In this section we describe the general modeling framework given the two sources of data. 
We develop a hierarchical model that links the true, unobserved BCS for each individual to two imperfect observation sources --- manual ordinal annotations and AI-derived compositional confidences --- while simultaneously relating the latent score to ecological covariates. 
The framework accounts for annotation error, annotator heterogeneity, AI uncertainty, and the ordinal structure of the biological process. 
A basic layout of the framework is given in Figure~\ref{fig:ModelFramework}. 
Refer to Table~1 in Section~S.1 of the Supplementary Material \citep{Supplement} for a description of all data and parameters.

Let $N$ be the number of sequences, where each sequence contains one individual. 
We assume that each individual has a true latent ordinal BCS which represents the health of the individual. 
For sequence $i = 1, \dots, N$, let $r_i>0$ denote the number of images and define $R = \sum_{i=1}^N r_i$ as the total number of images in the data. 
Define $\Lambda^Z_i$ and $\Lambda^C_i$ as the set of indices within sequence $i$ that are manually annotated and scored by AI, respectively. 
Note that the length of the union of these two sets is equal to $r_i$, either could be empty, and their intersection need not be empty. 
This implies that all images produce at least one form of observation and some images can have both annotated and AI observations.

Let $L$ denote the number of ordered categories of BCS, where $L\geq2$. For manual annotations, define $Z_{i,k} \in \{1, \dots, L\}$ for image $k \in \Lambda^Z_i$ of sequence $i$. 
Let $\mathbf C_{i,k}=\left(C_{i,k,1}, \dots, C_{i,k,L} \right)^\top$ be the $L$-length compositional AI confidence for image $k \in \Lambda^C_i$ of sequence $i$. 
We say that an $L$-length vector is compositional if it belongs to the $(L-1)$-dimensional simplex (i.e., all elements are nonnegative and sum to 1).

The observed manual annotations and AI confidences are modeled as function of the true BCS. 
Let $Y_i \in \{1, \dots, L\}$ for $i = 1, \dots, N$ denote the true latent BCS for individual $i$. 
For the ordinal manual annotations, we assume a probit link function and model $Z_{i,k}$ 
\begin{equation}\label{eq:fZ}
\begin{aligned}
    f_{Z}(z | Y_i, \boldsymbol \nu_{Y_i}, \boldsymbol \phi_{Y_i}) &= \Prob(Z_{i,k}=z | Y_i, \boldsymbol \nu_{Y_i}, \boldsymbol \phi_{Y_i}) \\
    &= \int_{\phi_{Y_i, z-1}}^{\phi_{Y_i, z}} \frac{1}{\sqrt{2 \pi}} \exp \left[-\frac{1}{2} \left( t - \nu_{Y_i}^{(a_{i,k})} \right)^2 \right] dt \\
    &= \Phi \left(\phi_{Y_i, z} - \nu_{Y_i}^{(a_{i,k})} \right) - \Phi \left( \phi_{Y_i, z-1} - \nu_{Y_i}^{(a_{i,k})} \right)
\end{aligned}
\end{equation}
where $\Phi$ is the cumulative distribution function for a standard normal random variable.
Here, $a_{i,k}$ indexes the annotator for image $k$ of sequence $i$. The annotation probabilities are controlled by annotator- and BCS-specific means $\nu_{Y_i}^{(a_{i,k})}$ where $\boldsymbol \nu_{Y_i} = \{\nu_{Y_i}^{(b)}: b=1, \dots, A\}$ and $L+1$-length vector of cutoffs $\boldsymbol{\phi}_{y}$ where $-\infty = \phi_{y,0} < \phi_{y,1} < \dots < \phi_{y,L-1} < \phi_{y,L} = \infty$ for $y=1, \dots, L$. 
Annotator-specific parameters allow for variation in accuracy across annotators which is believed to be present in the data. We assume independence across scores within annotators, although this assumption could be relaxed to incorporate dependence if warranted. 

The AI derived compositional data, $\mathbf C_{i,k}$, are modeled using independent Dirichlet distributions with mean $\boldsymbol \alpha_y$ for true BCS $y$ and precision $\exp[ \omega_0 + \mathbf U_{i,k}^\top \boldsymbol \omega ]$, where $\mathbf U_{i,k}$ is a $q$-length vector of possible image-quality covariates for image $k$ of sequence $i$ and $\omega_0$ and $\boldsymbol \omega = (\omega_1, \dots, \omega_q)^\top$ are coefficients.
For each true BCS $y \in \{ 1, \dots, L \}$, $\boldsymbol \alpha_y = (\alpha_{y,1}, \dots, \alpha_{y,L})^\top$. 
Therefore the probability distribution function for the compositional AI responses for image $k \in \boldsymbol \Lambda^C_i$ of sequence $i$ is 
\begin{equation} \label{eq:fC}
    f_{\mathbf C}(c_1, \dots, c_L | Y_i, \boldsymbol \alpha_{Y_i}, \omega_0, \boldsymbol \omega) 
    = \left( \frac{\Gamma \left( \exp[ \omega_0 + \mathbf U_{i,k}^\top \boldsymbol \omega ] \right)}{\prod_{\ell=1}^L \Gamma(\exp[ \omega_0 + \mathbf U_{i,k}^\top \boldsymbol \omega ] \boldsymbol \alpha_{Y_i, \ell})} \right) \prod_{\ell=1}^L c_{\ell}^{\exp[ \omega_0 + \mathbf U_{i,k}^\top \boldsymbol \omega ] \boldsymbol \alpha_{Y_i,\ell}-1}.
\end{equation}
We restrict $\boldsymbol \alpha_\ell$ to the $(L-1)$-dimensional simplex so that it is a valid mean for $\mathbf C_{i,k}$. 
The precision controls how tightly the values cluster around the mean, where higher precisions correspond to more consistent AI confidences within the sequence. 

Finally, we turn to modeling the latent process, $Y_i$, which captures the relationship between true BCS and possible individual level and environmental covariates.
Given that $Y_i$ is an ordinal variable, we again employ a probit link function. 
Probit models for ordinal random variables have been used to study various types of environmental processes \citep{Schliep2013, Schliep2015, Schliep2018} as they offer efficient computation within the Bayesian hierarchical model framework.
We define the probit regression model as
\begin{equation} \label{fY}
\begin{aligned}
    f_{Y}(y | \beta_0, \boldsymbol \beta, \boldsymbol \theta) &= \Prob(Y_i=y | \beta_0, \boldsymbol \beta, \boldsymbol \theta) \\
    &= \int_{\theta_{y-1}}^{\theta_{y}} \frac{1}{\sqrt{2 \pi}} \exp \left[-\frac{1}{2} \left( t - \beta_0 - \mathbf X_i^\top \boldsymbol \beta \right)^2 \right] dt \\
    &= \Phi(\theta_{y} - \beta_0 - \mathbf X_i^\top \boldsymbol \beta) - \Phi(\theta_{y-1} - \beta_0 - \mathbf X_i^\top \boldsymbol \beta)
\end{aligned}
\end{equation}
where $\mathbf X_i$ is a $p$-length vector of covariates for sequence $i$, $\beta_0$ and $\boldsymbol \beta = (\beta_1, \dots, \beta_p)^\top$ are the set of regression coefficients, and $L+1$-length vector $\boldsymbol \theta$ where $-\infty=\theta_0 < 0 = \theta_1 < \dots < \theta_{L-1} < \theta_L = \infty$ are the cutoffs for the ordinal values. 

\begin{figure}
    \centering
    \includegraphics[width=0.75\linewidth]{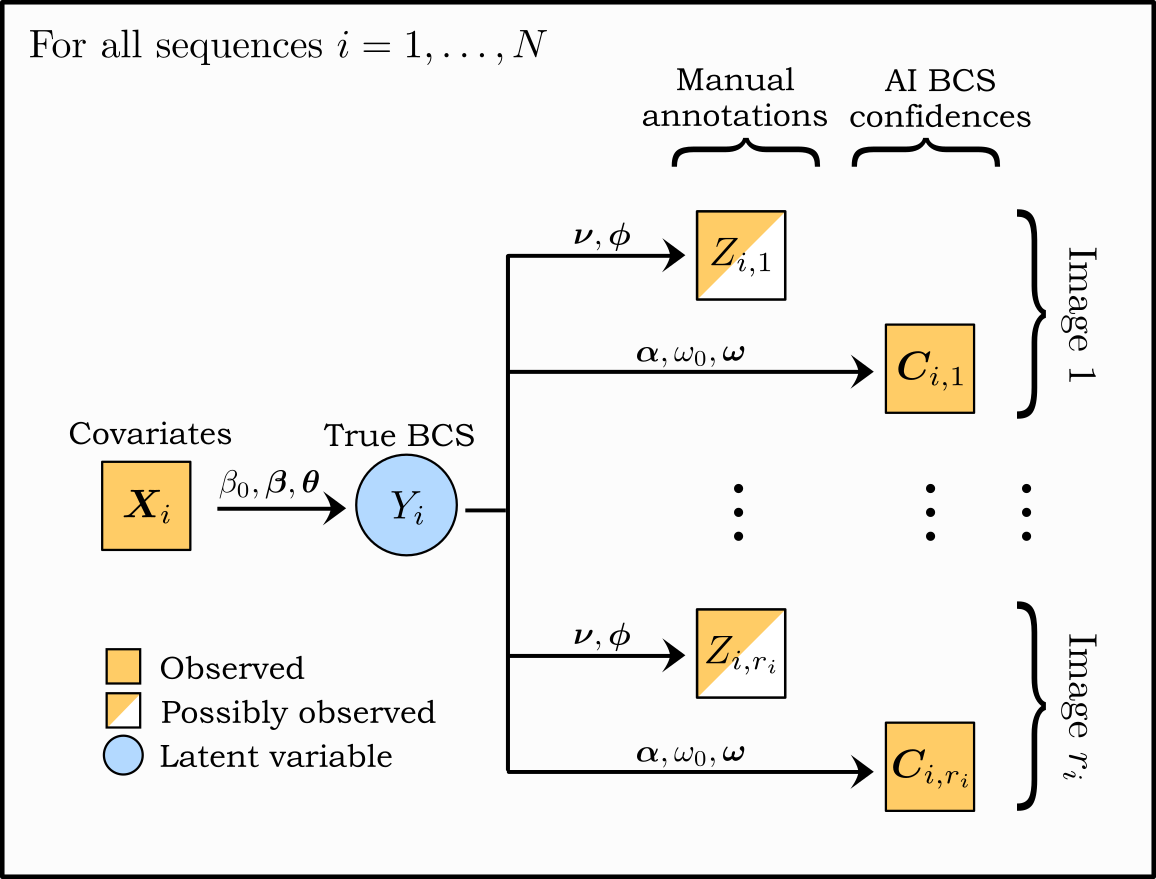}
    \caption{Visual representation of the model framework as described in Section~\ref{sec:Model.Framework}.}
    \label{fig:ModelFramework}
\end{figure}

\subsection{Model inference}

We specify our model hierarchically and obtain inference through a Bayesian framework. 
Prior distributions are assigned to all model parameters. 
We assign independent normal prior distributions with mean zero and variance $1$ to each coefficient parameter, $\beta_1, \dots, \beta_p$. 
Normal priors with mean $0$ and variance $10$ are assigned to $\beta_0$, $\omega_0$, and $\omega_1, \dots, \omega_q$. 
For parameter identifiability, we fix $\theta_1=\phi_{1,1}=\dots=\phi_{L,1}=0$ without loss of generality. 
For computational efficiency, the cut-off parameters are both reparameterized to alleviate the restricted distributions such that $\theta_y=\theta_{y-1} + e^{\tilde \theta_y}$ for $y = 2, \dots, L-1$ and $\phi_{y, z}=\phi_{y, z-1} + e^{\tilde \phi_{y, z}}$ for $y = 1, \dots, L$ and $z = 2, \dots, L-1$. 
The priors for these cutoff parameters are Log-Normally distributed with variance parameter $10$. 
The mean values for $\theta_2, \dots, \theta_{L-1}$ are equally spaced and chosen such that the \emph{a priori} probability that $Y_i$ is in any given category is at most $0.5$. 
The mean for each $\phi_{y,z}$ is such that annotators are assumed to be $95\%$ accurate, with equal probability given to all other categories. 
Because we assume the annotators are relatively consistent, we let the mean $\nu_y^{(b)}$ for each category $y$ and annotator $b$ be normally distributed centered around some latent $\tilde \nu_y$ with small variance equal to $0.2$. 
We let $\tilde \nu_y$ be uniformly distributed within its appropriate cutoffs $[\phi_{y, y-1}, \phi_{y, y}]$. 
We let $\boldsymbol \alpha_y$ be Dirichlet distributed where the $y$th element has mean $0.4$ and the remaining $0.6$ is split evenly across the other $L-1$ elements.
For example, if $L=5$ then $\boldsymbol \alpha_1 \sim \text{Dir}([0.4, 0.15, 0.15, 0.15, 0.15])$. 
A higher weight on the correct prediction provides model stability and is justified by validation of the AI model. Low relative precision and equal weight across remaining scores ensures the prior distribution remains relatively uninformative.

The posterior distribution is proportional to the product of the likelihood functions of the manual annotations and AI confidences, the process model for the latent true BCSs, and the prior distributions of the parameters. 
Define
$\boldsymbol \nu = \left \{\boldsymbol \nu_{y}: y = 1, \dots, L \right \}$,
$\boldsymbol {\tilde \nu} = \left \{ \tilde \nu_1, \dots, \tilde \nu_L \right \}$,
$\boldsymbol \phi = \left \{\boldsymbol \phi_{y}: y = 1, \dots, L\right\}$,
$\boldsymbol \alpha = \left \{ \boldsymbol \alpha_1, \dots, \boldsymbol \alpha_L \right \}$,
$\mathbf Z = \{ Z_i: i \in \boldsymbol \Lambda^Z_i \}$,
$\mathbf C = \{ \mathbf C_{i,k}: k = 1, \dots, r_i,i \in \boldsymbol \Lambda^C_i \}$, and $\mathbf Y = \{ Y_i: i=1, \dots, N \}$. 
The full posterior distribution can be written as
\begin{equation}
    \begin{aligned}
        &\left[ \boldsymbol \nu, \tilde{\boldsymbol \nu}, \boldsymbol \phi, \boldsymbol \alpha, \omega_0, \boldsymbol \omega, \beta_0, \boldsymbol \beta, \boldsymbol \theta, \mathbf Y | \mathbf Z, \mathbf C \right]
        \propto \\
        &\quad \left[ \mathbf Z | \mathbf Y, \boldsymbol \nu, \boldsymbol \phi \right] 
        \left[ \mathbf C | \mathbf Y, \boldsymbol \alpha, \omega_0, \boldsymbol \omega \right]
        \left[ \mathbf Y | \beta_0, \boldsymbol \beta, \boldsymbol \theta \right]
        \left[ \boldsymbol \nu | \tilde{\boldsymbol \nu}, \boldsymbol \phi \right]
        \left[ \tilde{\boldsymbol \nu} | \boldsymbol \phi \right]
        \left[ \boldsymbol \phi \right]
        \left[ \boldsymbol \alpha \right]
        \left[ \omega_0 \right]
        \left[ \boldsymbol \omega \right]
        \left[ \beta_0 \right]
        \left[ \boldsymbol \beta \right]
        \left[ \boldsymbol \theta \right].    
    \end{aligned}
\end{equation}

Posterior samples are obtained through a customized Markov chain Monte Carlo (MCMC) sampling algorithm. 
A Gibbs step is used to sample $\mathbf Y$ directly, while an adaptive Metropolis-Hastings algorithm is used for the remaining parameters. 
Details about sampling are given in the appendix.
To handle numerical issues with the Dirichlet distribution when an element of $\mathbf C_{i,k}$ is exactly zero or one, we adjust $\mathbf C_{i,k}$ by first choosing $\zeta=1e-12$ and $\zeta=1e-3$ for the simulation study and case study, respectively. 
Then, for all $\mathbf C_{i,k}$ which contain any elements smaller than $\zeta$, $\zeta$ is added to each element and divided by the sum of the vector to preserve the unitary sum. 
This is done for numerical stability of the algorithm and results were found to be robust to this adjustment.

\section{Simulation study} \label{sec:Simulation}

The simulation study evaluates how well the full data-fusion model recovers environmental effects and predicts latent BCS relative to simpler modeling shortcuts commonly used in camera-trap AI applications.

\subsection{Model settings}
The following simulation was conducted to evaluate the proposed modeling approach for the two data sources. 
To this end, we chose $5$ relevant model scenarios, listed below, for comparison.

\emph{Scenario 1 (linear)} assumes a linear regression model and is one of the simplest approaches one might take to model the data. 
Under this approach, the ordinal nature of the observed data are ignored and instead, continuous response variables $\tilde Y_i$ for $i = 1, \dots, N$ are extracted.
If sequence $i$ has annotated images, then all AI confidences are ignored and $\tilde Y_i$ is the mean of the annotations.
Otherwise, $\tilde Y_i$ is the mean of the weighted average of AI confidences for sequence $i$. 
All confidences for which the maximum value is less than some prespecified threshold $T$ are discarded prior to computing $\tilde Y_i$ to reduce uncertainty following, for example, \cite{Cowans2024} and \cite{Ringwaldt2025}.
We model these continuous response variables using a Bayesian linear regression model to enable easy comparison.

\emph{Scenario 2 (maximum)} applies a threshold to the AI confidences to filter out less confident predictions.
The manual annotations are ignored. The ordinal value with the maximum confidence is extracted and the median of these values across all images within the sequence is retained as the observed ordinal value for individual $i$. 
This is a common interpretation of AI confidences.  
For simplicity, the model in this scenario assumes that $Y_i$ can be observed without error.
As with scenario 1, AI confidences with maximum value less than $T$ are discarded.

Scenarios 3-4 use the proposed model from Section~\ref{sec:Model} but are limited to a single data source. \emph{Scenario 3 (ordinal-only)} omits the AI confidences while \emph{Scenario 4 (compositional-only)} omits the manual annotations.
\emph{Scenario 5 (full model)} uses all available data under the full proposed model in Section 3.
Comparisons between the models under scenarios 3 and 4 to the full model (scenario 5) will highlight the benefits of each data source with regard to model inference and prediction.

\subsection{Model comparison}
We assess the 5 models based on their ability to address the specific goals of the analysis. 
First, to assess how well the model estimates the coefficients for the environmental drivers, the mean squared error (MSE) and empirical coverage for each $\boldsymbol \beta$ are reported under each model setting.
Second, to assess the model fit with respect to the true BCS values, $Y_1, \dots, Y_N$, the ranked probability score (RPS) is calculated with respect to the posterior distribution of $\mathbf Y$ given the data.
Third, to evaluate the predictive ability of the model, we compute the RPS for each out-of-sample individual at a new location (i.e., using the posterior prediction distribution of $\mathbf Y_0$ given the data, where $\mathbf Y_0$ is the vector of latent true ordinal BCSs for a set of out-of-sample individuals). 
RPS is a proper scoring rule \cite[][]{Gneiting2007} in that it is minimized by the true distribution and accounts for both accuracy and precision in prediction. The posterior predictive distribution of the ordinal score is compared to the true value of the ordinal data and lower values correspond to better predictions. 
Details for computing RPS under our model are given in Section~S.4 of the Supplementary Material \citep{Supplement}. 
Note that since BCS observations under scenario 1 are continuous, the RPS analysis is omitted due to lack of comparison.

\subsection{Simulation settings}

The simulation settings are chosen to mimic realistic ecological data from camera traps. 
We randomly generate 100 data sets and fit each of the 5 models to each dataset according to the preprocessing outlined in Section 4.1. 
For each simulation, we assume $N=500$ training sequences and $1000$ testing sequences. 
For sequence $i = 1, \dots, N$, the number of images in the sequence, $r_i$, is randomly assigned  a value in the set $\{1, 3, 5, 10\}$ with equal probability. 
These sequence frequencies are chosen to approximately represent the quantiles in the Candid Critters dataset.
Each image is assumed to have AI confidences whereas a pre-specified proportion of the images are randomly selected to have manual annotations. 
For scenarios 3 and 5, we randomly select $10\%$, $20\%$, and $50\%$ of the images to manually annotate, and for scenario 1 we randomly select $20\%$ of the images to manually annotate.
We assume $L=5$ ordered categories for the latent ordinal process of interest. 
We assume $p=6$ and set $\beta_0=1.38$ and$\boldsymbol \beta=(0.2, -0.3,  0.0,  0.0,  0.2, -0.3)^\top$. Letting $q=1$, we set $\omega_0=0$ and $\omega=1$.
We fix $\boldsymbol \theta$ so that the maximum probability of $Y_i$ being in categories 2, 3, and 4 are $0.2$, $0.5$, and $0.3$, respectively. 
Similarly, we assume there are $3$ annotators and specify $\tilde \nu$ and $\boldsymbol \phi$ such that an annotator with $\boldsymbol \nu^{(a)} = \tilde{\boldsymbol \nu}$ would be $95\%$ accurate, and randomly generate each $\boldsymbol \nu^{(a)} \sim \text{MVN}(\tilde{\boldsymbol \nu}, 0.1 \mathbf I_5)$, where $\mathbf I_5$ is the $5 \times 5$ identity matrix for $a=1,2,3$. 
We set each $\boldsymbol \alpha_y$ such that the AI is expected to give $60\%$ confidence for the true value $Y_i=y$ and $10\%$ confidence for all other categories. For example, $\boldsymbol \alpha_1=(0.6,0.1,0.1,0.1,0.1)^\top$.
For each sequence $i$, we randomly sample a row from the covariates matrices, $\mathbf X_i$ and $\mathbf U_i$, where $p=6$ and $q=1$. 
We assume $\zeta = 1e-12$ for numerical stability. 
For scenarios 1 and 2 we choose thresholds of $T=0.75$, $0.9$, and $0.99$ (i.e., at least $75\%$, $90\%$, and $99\%$ confident, respectively).

\subsection{Results}
We fit each model by running the MCMC sampling algorithm for $15000$ iterations, of which the first $10000$ are discarded as burn-in. The remaining $5000$ posterior samples are used for inference and prediction.
Convergence was assessed by visual inspection of trace plots and no issues of convergence were detected.

Recall that each model was trained using a different subset of the simulated dataset. 
For example, the maximum model thresholded at $90\%$ uses only the highest confidence image in each sequence and discards sequences with no confidence of at least $90\%$; the ordinal-only model only trains on annotated images. 
See Table~2 in Section~S.2 of the Supplementary Material \citep{Supplement} for more details on the number of images and sequences used in each scenario.
The models are assessed using identical datasets for out-of-sample validation, enabling direct comparisons across models.
Therefore direct comparisons can be made for out-of-sample results but not in-sample results.

We present MSE for each regression coefficient in Table~\ref{tab:MSE}. 
The full model had the lowest MSE for all non-zero coefficients, while the linear model had slightly lower MSE for the coefficients which were equal to zero.
Coverage for $95\%$ credible intervals (CIs) for each coefficient are given in Table~\ref{tab:coverage}. 
Approximately $95\%$ coverage or higher is achieved by the full, ordinal-only, and compositional-only models. 
The linear and maximum models exhibit under-coverage.
Detection, which we define as the proportion of $95\%$ CIs correctly not containing zero for coefficients which are not equal to zero and correctly containing zero for coefficients which equal zero, are given in Table~\ref{tab:detection}. 
The full and compositional-only models achieve approximately $95\%$ detection or higher, while the other models exhibit under-detection, especially the maximum models and ordinal-only with $10\%$-$20\%$ of images annotated. 

\begin{table}[h!]
    \centering
    \begin{tabular}{l c c c c c c} \hline Scenario & $\beta_{1}=0.2$ & $\beta_{2}=-0.3$ & $\beta_{3}=0.0$ & $\beta_{4}=0.0$ & $\beta_{5}=0.2$ & $\beta_{6}=-0.3$  \\ \hline (1) Linear & & & & & & \\ \quad Thresholded at $75\%$ & $5.81$ & $7.46$ & $\mathbf{1.89}$ & $\mathbf{2.75}$ & $4.17$ & $6.17$ \\ \quad Thresholded at $90\%$ & $5.82$ & $6.52$ & $2.06$ & $2.91$ & $3.70$ & $5.49$ \\ \quad Thresholded at $99\%$ & $5.06$ & $5.98$ & $2.61$ & $3.00$ & $3.76$ & $5.29$ \\ (2) Maximum & & & & & & \\ \quad Thresholded at $75\%$ & $5.96$ & $8.01$ & $3.17$ & $4.06$ & $5.48$ & $6.58$ \\ \quad Thresholded at $90\%$ & $6.86$ & $8.58$ & $4.05$ & $5.01$ & $5.51$ & $8.02$ \\ \quad Thresholded at $99\%$ & $7.85$ & $12.40$ & $9.13$ & $7.20$ & $6.50$ & $11.69$ \\ (3) Ordinal-only & & & & & & \\ \quad $10\%$ annotated & $10.32$ & $10.92$ & $9.46$ & $13.02$ & $9.60$ & $13.97$ \\ \quad $20\%$ annotated & $5.97$ & $5.71$ & $5.48$ & $7.43$ & $5.80$ & $6.84$ \\ \quad $50\%$ annotated & $3.40$ & $3.55$ & $3.13$ & $4.28$ & $4.03$ & $4.61$ \\ (4) Compositional-only & $2.55$ & $3.32$ & $2.61$ & $3.38$ & $3.30$ & $4.37$ \\ (5) Full model & & & & & & \\ \quad $10\%$ annotated & $\mathbf{2.49}$ & $3.16$ & $2.44$ & $3.33$ & $3.28$ & $\mathbf{4.16}$ \\ \quad $20\%$ annotated & $2.63$ & $2.97$ & $2.26$ & $3.24$ & $3.27$ & $4.22$ \\ \quad $50\%$ annotated & $2.55$ & $\mathbf{2.92}$ & $2.32$ & $2.99$ & $\mathbf{3.03}$ & $4.20$ \\ \hline \end{tabular}
    \caption{Mean squared error for each $\beta$ coefficient for each of the models in scenarios 1 - 5. True values for each coefficient are given across the top to identify significant and non-significant effects. Mean squared errors are reported by $10^{-3}$.}
    \label{tab:MSE}
\end{table}

\begin{table}[h!]
    \centering
    \begin{tabular}{l c c c c c c} \hline Scenario & $\beta_{1}=0.2$ & $\beta_{2}=-0.3$ & $\beta_{3}=0.0$ & $\beta_{4}=0.0$ & $\beta_{5}=0.2$ & $\beta_{6}=-0.3$  \\ \hline (1) Linear & & & & & & \\ \quad Thresholded at $75\%$ & $0.76$ & $0.71$ & $0.96$ & $0.96$ & $0.86$ & $0.74$ \\ \quad Thresholded at $90\%$ & $0.78$ & $0.83$ & $0.97$ & $0.98$ & $0.90$ & $0.81$ \\ \quad Thresholded at $99\%$ & $0.89$ & $0.87$ & $0.96$ & $0.99$ & $0.90$ & $0.90$ \\ (2) Maximum & & & & & & \\ \quad Thresholded at $75\%$ & $0.80$ & $0.78$ & $0.95$ & $0.95$ & $0.82$ & $0.84$ \\ \quad Thresholded at $90\%$ & $0.81$ & $0.84$ & $0.98$ & $0.93$ & $0.88$ & $0.86$ \\ \quad Thresholded at $99\%$ & $0.84$ & $0.84$ & $0.94$ & $0.97$ & $0.92$ & $0.85$ \\ (3) Ordinal-only & & & & & & \\ \quad $10\%$ annotated & $0.93$ & $0.97$ & $0.97$ & $0.97$ & $0.97$ & $0.94$ \\ \quad $20\%$ annotated & $0.94$ & $0.96$ & $0.97$ & $0.93$ & $0.96$ & $0.91$ \\ \quad $50\%$ annotated & $0.96$ & $0.93$ & $0.94$ & $0.96$ & $0.91$ & $0.91$ \\ (4) Compositional-only & $0.97$ & $0.97$ & $0.98$ & $0.97$ & $0.94$ & $0.90$ \\ (5) Full model & & & & & & \\ \quad $10\%$ annotated & $0.97$ & $0.97$ & $0.99$ & $0.96$ & $0.94$ & $0.91$ \\ \quad $20\%$ annotated & $0.97$ & $0.99$ & $0.98$ & $0.96$ & $0.94$ & $0.93$ \\ \quad $50\%$ annotated & $0.98$ & $0.97$ & $0.97$ & $0.97$ & $0.93$ & $0.90$ \\ \hline \end{tabular}
    \caption{Empirical coverage based on $95\%$ credible intervals for each $\beta$ coefficient for model scenarios 1 - 5. }
    \label{tab:coverage}
\end{table}

\begin{table}[h!]
    \centering
    \begin{tabular}{l c c c c c c} \hline Scenario & $\beta_{1}=0.2$ & $\beta_{2}=-0.3$ & $\beta_{5}=0.2$ & $\beta_{6}=-0.3$ & $\beta_{3}=0.0$ & $\beta_{4}=0.0$  \\ \hline (1) Linear & & & & & & \\ \quad Thresholded at $75\%$ & $0.80$ & $1.00$ & $0.92$ & $1.00$ & $0.96$ & $0.96$ \\ \quad Thresholded at $90\%$ & $0.79$ & $1.00$ & $0.91$ & $1.00$ & $0.97$ & $0.98$ \\ \quad Thresholded at $99\%$ & $0.78$ & $0.99$ & $0.90$ & $0.99$ & $0.96$ & $0.99$ \\ (2) Maximum & & & & & & \\ \quad Thresholded at $75\%$ & $0.74$ & $0.98$ & $0.78$ & $0.99$ & $0.95$ & $0.95$ \\ \quad Thresholded at $90\%$ & $0.69$ & $0.95$ & $0.79$ & $0.98$ & $0.98$ & $0.93$ \\ \quad Thresholded at $99\%$ & $0.60$ & $0.82$ & $0.64$ & $0.83$ & $0.94$ & $0.97$ \\ (3) Ordinal-only & & & & & & \\ \quad $10\%$ annotated & $0.51$ & $0.80$ & $0.65$ & $0.90$ & $0.97$ & $0.97$ \\ \quad $20\%$ annotated & $0.74$ & $0.98$ & $0.86$ & $0.99$ & $0.97$ & $0.93$ \\ \quad $50\%$ annotated & $0.91$ & $1.00$ & $0.93$ & $1.00$ & $0.94$ & $0.96$ \\ (4) Compositional-only & $0.94$ & $1.00$ & $0.95$ & $1.00$ & $0.98$ & $0.97$ \\ (5) Full model & & & & & & \\ \quad $10\%$ annotated & $0.98$ & $1.00$ & $0.98$ & $1.00$ & $0.99$ & $0.96$ \\ \quad $20\%$ annotated & $0.96$ & $1.00$ & $0.97$ & $1.00$ & $0.98$ & $0.96$ \\ \quad $50\%$ annotated & $0.99$ & $1.00$ & $0.96$ & $1.00$ & $0.97$ & $0.97$ \\ \hline \end{tabular}
    \caption{Proportion of $95\%$ credible intervals that correctly identify significant (do not cover zero for $\beta \neq 0$) and non-significant (do cover zero for $\beta = 0$) effects for each model scenario 1 - 5.}
    \label{tab:detection}
\end{table}

\begin{figure}
    \centering
    \includegraphics[width=\linewidth]{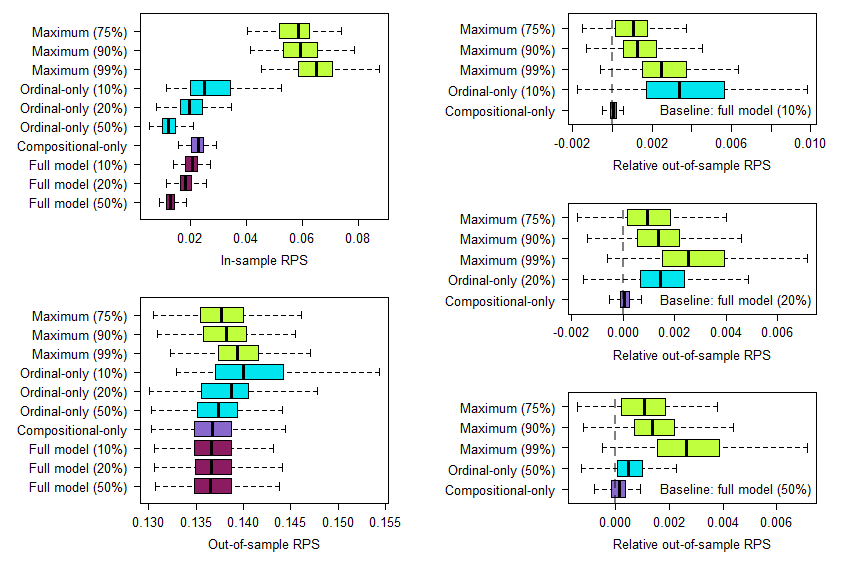}
    \caption{Ranked probability score (RPS) for in-sample model validation and out-of-sample model evaluation (left) and relative out-of-sample RPS with respect to the full model (right) for simulated data. The full and ordinal-only models were fit using different percentages of annotated images and the maximum model was fit with different thresholds. The linear scenario is not included due to incompatibility of the RPS metric.}
    \label{fig:RPS_plot}
\end{figure}

A summary of the RPS results for in-sample model validation are given in Figure~\ref{fig:RPS_plot}. 
Recall that RPS rewards both accuracy and precision.
Since the maximum model removes images below a given threshold, increasing the threshold results in higher quality yet fewer images used in training the model.  
As such, we find that increasing the threshold does not improve model performance.
For the ordinal-only model, as more images are annotated, model performance improves. 
Similar performance gains are seen with the full model as more images are annotated. 
Importantly, the full model out-performs the compositional-only model at each level, and significantly so when 50\% of images are annotated.
Note that for in-sample performance, the linear models, the maximum models, and the ordinal-only models are only validated on the subset of data used in fitting the respective models.
The compositional and full models are validated on the full data, which includes those images that have compositional-only data, which are in general more variable and more challenging to predict. 
As such, these gains in in-sample RPS for the full model are significant in relation to the ordinal-only model.

A summary of the RPS results for out-of-sample model validation are also shown in Figure~\ref{fig:RPS_plot}. 
Out-of-sample performance is being measured using identical validation datasets, making the loss in performance much more apparent for models with reduced data.
In particular, the ordinal-only models perform worse than the compositional-only and full models.

The three panels on the right-hand side of Figure~\ref{fig:RPS_plot} show a more direct comparison of the out-of-sample performance.
To reduce the effect of variation in the simulated datasets, we compare each model to the full model under different levels of annotation. 
Shown in each boxplot is the difference in RPS values of each model considered and the baseline full model. 
We find that no model outperforms the full model. As the number of annotations increases, the ordinal-only model approaches the full model. 
The full model relative to the compositional-only model also shows improved predictive performance as the number of annotations used in the full model increases. 

Across all simulation scenarios, the full model is competitive with respect to each metric considered, which is not achieved by any of the other models considered. 
The compositional-only model performed very similarly to the full model, but was outperformed with respect to in-sample RPS. 
We find minimal gains from including the manual annotations under these settings in terms of prediction, although they still improved uncertainty quantification.
Therefore the full model as proposed in Section~\ref{sec:Model} has the best performance with respect to estimating the effects of the environmental drivers --- including recovering the true values, achieving valid coverage, and detecting differences from zero --- and prediction.

\section{Case study} \label{sec:CaseStudy}

\begin{figure}[ht]
    \centering
    \includegraphics[width=1\linewidth]{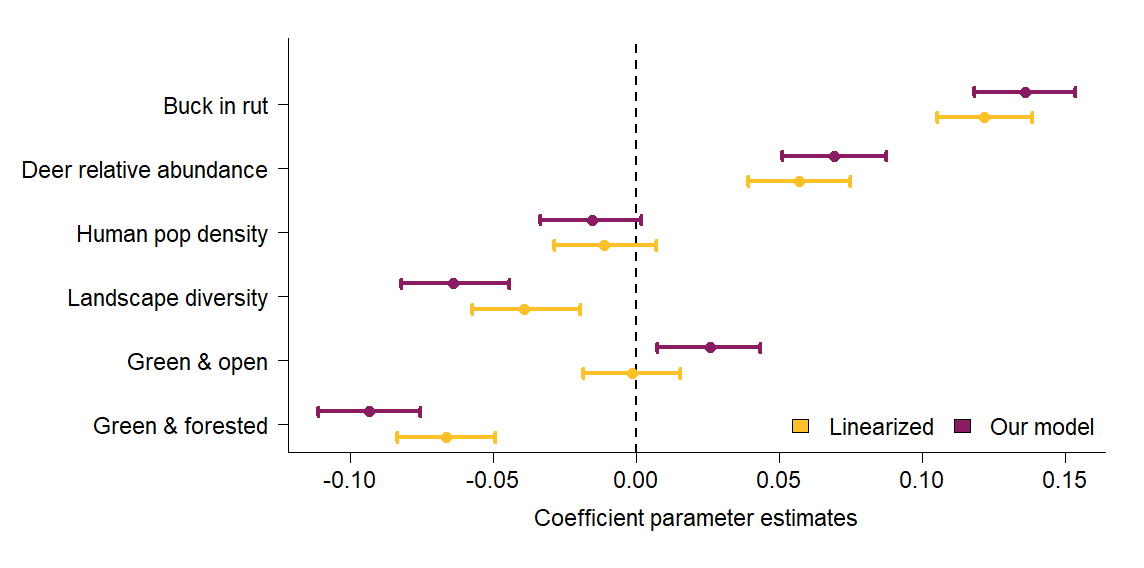}
    \caption{Posterior means and $95\%$ credible intervals for each regression coefficient, $\beta$, for the Candid Critters analysis under our model compared to the linearized model.}
    \label{fig:beta_posteriors}
\end{figure}

We fit the proposed model to the Candid Critters camera trap dataset. 
For comparison, we also fit the linear model thresholded at $90\%$ per expert recommendation. 
This choice results in $30\%$ of the sequences being discarded. 
For sequences with more than $30$ images, we randomly select a subset of $30$ images to limit the computational load.
We ran each model for $15000$ MCMC iterations, the first $10000$ of which are discarded as burn-in and confirmed convergence of the chains by visual inspection of the traceplots of the posterior samples. 
The $95\%$ CIs for the regression coefficients are given in Figure~\ref{fig:beta_posteriors}. 
The coefficient with the largest magnitude corresponds to the indicator variable for whether the deer is a buck in rut. 
Because bucks tend to gain weight to compete for does during rut, this very strong positive correlation with BCS is expected. 
We also expect that green, open habitats provide an ideal food source for deer, and so should correspond to a higher BCS \cite[][]{Jones2010}. 
Importantly, this effect is detected by the full model but not by the linear model.

The effects of the remaining covariates on deer health are not well established in ecological literature and therefore provide novel inference. 
Our model finds that deer relative abundance is positively correlated with BCS, while landscape diversity, green forested habitat, and human population density are negatively correlated.
The difference in effects between green and open versus forested habitats likely results from the ability of the deer to access the new vegetation growth.

\begin{figure}[ht]
    \centering
    \includegraphics[width=1\linewidth]{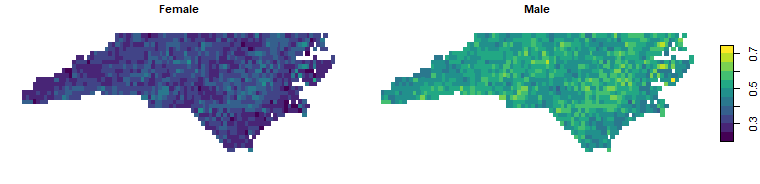}
    \caption{Spatial variation in predictions of probability of having a high (4 or 5) body score for females (left) and males (right) in November 2018.}
    \label{fig:prop_high_maps}
\end{figure}

Predictions are made for a grid of $1264$ sites over North Carolina. 
The NDVI values used to calculate ``green and open'' and ``green and forested'' in these predictions are drawn from November 2018 (see Figure~\ref{fig:covariate_maps}). 
The predicted probability of having a high (4 or 5) body condition for male (i.e. bucks in rut) and female deer are compared in Figure~\ref{fig:prop_high_maps}. 
The male deer are predicted to have a much higher BCS overall, and body condition tends to be lower near the mountains and coast. 

\begin{figure}
    \centering
    \includegraphics[width=0.39\linewidth]{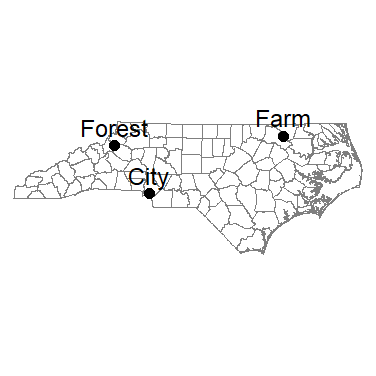}
    \includegraphics[width=0.6\linewidth]{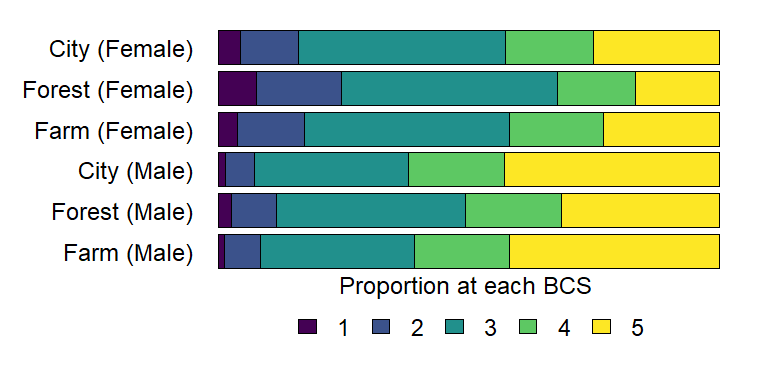}
    \caption{(Left) Sites where covariate values were drawn are representative of city, forest, and farm locations. (Right) Posterior predictions of body condition distributions for selected sites for male and female deer during the fall season.}
    \label{fig:site_prop}
\end{figure}

We also make predictions at several sites of interest which are representative of city, forest, and farm habitats. 
The covariate values for these illustrative sites are given in Table~\ref{tab:site_covariates} and the locations of the sites are shown across the state in Figure~\ref{fig:site_prop}. 
Also shown in Figure ~\ref{fig:site_prop} is the predicted probability of an individual deer being in each of the given BCS categories. 
Male deer are predicted to have higher body condition than female deer during the fall season.
Cities and farms have similar results, while forests have a lower proportion of deer with high body condition. 
This is consistent with the intuition that farmland provides good habitat for deer with a large amount of available vegetation at close proximity to forest habitat.

\begin{table}[]
    \centering
    \begin{tabular}[h]{r c c c c c}
        \hline
          & Deer relative abundance & Human pop density & Landscape diversity & Green \& open & Green \& forested\\
        \hline
        City & 0.02 & 2043.00 & 0.14 & -0.52 & -2.55\\
        Forest & 0.02 & 69.70 & 0.00 & -0.88 & 0.33\\
        Farm & 0.02 & 72.30 & 0.03 & 0.17 & -1.75\\
        \hline
    \end{tabular}
    \caption{Environmental covariate values for representative sites of a city, forest, and farm location.}
    \label{tab:site_covariates}
\end{table}

\section{Discussion} \label{sec:Discussion}

We proposed a novel Bayesian hierarchical framework for the integration of manual annotations and AI confidences from camera trap data.
Current approaches often under-utilize or simplify the available data. 
In particular, they often fail to fully integrate the manual annotations and AI confidences, lack proper uncertainty quantification, and may discard large amounts of data.
By jointly modeling manual annotations and AI confidences, we leverage the strengths of both data sources by using the manual annotations to provide calibration and uncertainty quantification, while the AI confidences provide large quantities of inexpensive data spanning broader spatial and environmental extents.

The proposed model is fitted to the Candid Critters camera trap dataset to provide ecological inference with respect to deer BCS and environmental drivers across North Carolina. 
A simplified linear model was also fitted for comparison. 
Both models suggest a significant increase in BCS for bucks in rut, as is expected from basic deer biology.  
The proposed model finds that BCS is positively correlated with green open habitat, while the linear model does not detect a significant relationship. 
Our model also provides novel inference, suggesting that BCS is positively correlated with deer relative abundance and negatively correlated with landscape diversity, green forested habitat, and possibly human population density.
Future work looks to expand our analysis to Snapshot USA data, a larger camera trap study which spans the United States \cite[][]{Rooney2025}. 
This would enable inference on the effect of other important health indices, such as Chronic Wasting Disease, a prion disease which has been spreading across the country and internationally \cite[][]{CWD}.

Utilizing AI-integrated camera trap data opens up new areas of exploration in ecology.
Body condition in deer is a mostly unexplored area of research.
The existing studies are limited by the cost of data collection.
For example, \cite{Simard2014} conduct a similar analysis of the relationship between environmental drivers and deer body condition in the fall.
They consider deer collected by hunters, leading to a much smaller and possibly biased sample. 
Our results are similar in finding that body condition is lower in female deer, although they find that body condition is lower in areas with high deer density, whereas we found a positive correlation. 

The benefit of the manual annotations is that they enable calibration of the joint model, a critical step in ensuring valid inference.
However, one limitation we found in our analysis using the proposed model is that a high proportion of manual annotations are necessary.
In fact, the compositional-only model was shown to be highly competitive with respect to the other simplified settings. 
This result may not be robust to other camera trap analyses and should be explored in future studies with different response types, such as disease conditions.
We predict that the effect of the manual annotations depends in part on how well the AI model is calibrated.

Another limitation of using AI predictions is the inability to account for sequences containing multiple deer.
This is due to the difficulty associated with identifying and tracking individual deer between images.
Models that can incorporate multiple animals are particularly important for species that commonly travel in groups, such as white-tailed deer, elk, and wild boar.
Removing sequences containing multiple animals could introduce bias if group size is correlated with health.
A potential solution would be to randomly select one image from each sequence containing multiple deer while retaining the entire sequence when there is only one deer present.
This would meet model assumptions by ensuring there is only one deer per sequence, mitigating possible bias.
However, using all images from all sequences is still an open research question. 

More extensive methodological development entails extending the approach to occupancy or abundance modeling which will focus on species identification. 
The branching taxonomic structure defining similarity between species requires statistical novelty beyond the ordinal latent variable modeling for BCS. 
In particular, we expect that AI models trained to identify species will have a harder time distinguishing closely related species than distantly related species.
It is also of interest to perform inference on multiple taxonomic levels, such as species or genus level, although current models are limited to one level.
Manual annotations may be more valuable for this type of model, and sequential design could be used to guide such annotation schemes.
Sequential design or active learning approaches inform model design by optimizing the labeling of data (e.g., the species identity) with respect to both model inference and uniformly sampling the available data.
This is an area of active research and has not yet been applied to occupancy and abundance models.

\section*{Significance statement}


Effectively monitoring species is a necessary step in conservation.
The widespread adoption of camera traps has greatly increased the scale at which surveys can be conducted, but the large number of images has caused a bottleneck during image processing.
Automatic classification using artificial intelligence (AI) presents a solution to this problem, however, applying statistical methods that are not equipped to handle AI model output can lead to misinformed management decisions.
We present a novel statistical method that effectively handles this new type of data to study deer body condition, a measure of deer health.
This method is applicable to any data of this structure, such as species identifications or classifications of medical images.

\section*{Data availability} \label{sec:DataAvailability}

The code for fitting the model and simulating data is available on GitHub at \url{https://github.com/acohen6/cohen2026_ordinal_fusion}.

\section*{Acknowledgments}
The first author was supported in part by the NSF-RTG: Uncertainty Quantification in Life Sciences (UQ4Life) (DMS-2342344) and Summer One Health Internship on Wildlife Health offered by the North Carolina Museum of Natural Sciences in collaboration with the North Carolina State Global One Health Academy. 
The manual annotations, including for use in training the AI model, were in part provided by the Seed Funding Grant, NCSU Office of University Interdisciplinary Programs.

\bibliography{bibliography}       

\newpage

\section*{Supplementary material}

\renewcommand{\thesection}{\Alph{section}}
\setcounter{section}{19}

\subsection{Notation} \label{sec:Supp.Notation}

Model notation is organized in Table~\ref{tab:notation}.
\begin{table}[h!]
    \centering
    \singlespacing
    \small
    \begin{tabular}{ c p{4cm} p{8cm} }
     \hline
     \textbf{Parameter} & \textbf{Dimensions} & \textbf{Meaning} \\
     \hline \hline
     $N$ & positive integers & number of sequences \\
     $R$ & positive integers & number of images \\
     $r_i$ & integer & number of images in sequence $i$ \\
     $\boldsymbol \Lambda^Z_i$ & sets up to size $r_i$ & indices of the images in sequence $i$ with manual annotations  \\
    $\boldsymbol \Lambda^C_i$ & sets up to size $r_i$ & indices of the images in sequence $i$ with AI confidences \\
     $L$ & positive integer & number of categories \\
     $p$ & positive integers & number of environmental covariates \\
     $q$ & positive integers & number of image-quality covariates \\
     $A$ & positive integer & number of annotators \\
     \hline \hline
     $Y_i$ & integers in the set $\{1, \dots, L\}$ & true latent BCS for sequence $i$ \\
     $Z_{i,k}$ &  integers in the set $\{1, \dots, L\}$ & ordinal BCS annotation for image $k$ of sequence $i$ \\
     $\mathbf C_{i,k}$ & $L$-length vector & compositional AI confidences for image $k$ of sequence $i$ \\
     \hline \hline
     $\mathbf X_i$ & $p$-length vector & environmental covariates for sequence $i$ \\
     $\mathbf U_{i,k}$ & $q$-length vector & image-quality covariates for image $k$ of sequence $i$ \\
     $a_{i,k}$ & integer & identity of the annotator for image $k$ of sequence $i$ \\
     \hline \hline 
     $\beta_0$ & real number & intercept coefficient for true latent BCS \\
     $\boldsymbol \beta$ & $p$-length vector & coefficients mapping environmental covariates to true latent BCS \\
     $\boldsymbol \theta$ & ordered vector of length $L+1$ & cutoffs for $Y$; $-\infty = \theta_0 < 0 = \theta_1 < \dots < \theta_{L-1} < \theta_L = \infty$ \\
     $\tilde{\boldsymbol \theta}$ & $(L-2)$-length vector & reparameterization of $\boldsymbol \theta$; $\tilde \theta_y = \text{log}(\theta_y - \theta_{y-1})$ for $y \in \{ 2, \dots, L-1 \}$. \\  
     $\boldsymbol \phi_y$ & ordered vector of length $L+1$ & cutoffs for $Z_{i,k}$ when $Y_i=y$; $-\infty = \phi_{y, 0} < 0 = \phi_{y, 1} < \dots < \phi_{y, L-1} < \phi_{y, L} = \infty$  \\
     $\tilde{\boldsymbol \phi_y}$ & $(L-2)$-length vector & reparameterization of $\boldsymbol \phi_y$; $\tilde \phi_{y,z} = \text{log}(\phi_{y, z}-\phi_{y, z-1})$ for $z \in \{ 2, \dots, L-1 \}$ \\  
     $\nu_{y}^{(a)}$ & real number & mean of latent annotation normal random variable for annotator $a$ when $Y_i=y$ \\
     $\tilde \nu_y$ & real number & latent mean parameter for $\nu_y^{(1)}, \dots, \nu_y^{(A)}$ \\
     $\boldsymbol \alpha_y$ & $L$-length vector & means for compositional responses if $Y_i=y$; $\boldsymbol \alpha_y$ in $(L-1)$-simplex \\
     $\omega_0$ & real number & intercept for the logged precision of compositional responses \\
     $\boldsymbol \omega$ & $q$-length vector & image-quality coefficients for precision of compositional responses \\
     \hline
    \end{tabular}
    \caption{Table of model notation.}
    \label{tab:notation}
\end{table}

\subsection{Simulation data details} \label{sec:Data.Details}
The median number of images and sequences used for each scenario in the simulation study are reported in Table \ref{tab:sim_num_valid}.
\FloatBarrier
\begin{table}[h!]
    \centering
    \begin{tabular}{l c c} \hline Scenario & Images & Sequences  \\ \hline (1) Linear & & \\ \quad Thresholded at $75\%$ & $837$ & $446$ \\ \quad Thresholded at $90\%$ & $710$ & $403$ \\ \quad Thresholded at $99\%$ & $590$ & $354$ \\ (2) Maximum & & \\ \quad Thresholded at $75\%$ & $1392$ & $424$ \\ \quad Thresholded at $90\%$ & $930$ & $336$ \\ \quad Thresholded at $99\%$ & $496$ & $213$ \\ (3) Ordinal-only & & \\ \quad $10\%$ annotated & $238$ & $181$ \\ \quad $20\%$ annotated & $452$ & $272$ \\ \quad $50\%$ annotated & $1049$ & $400$ \\ (4) Compositional-only & $2376$ & $500$ \\ (5) Full model & & \\ \quad $10\%$ annotated & $2376$ & $500$ \\ \quad $20\%$ annotated & $2376$ & $500$ \\ \quad $50\%$ annotated & $2376$ & $500$ \\ \hline \end{tabular}
    \caption{Median number of images and sequences used for each scenario in the simulation study in Section~4. Note that for each of the 100 replications, the same dataset was used across all scenarios, but each scenario used a different subset of the simulated data.}
    \label{tab:sim_num_valid}
\end{table}

\FloatBarrier
\subsection{Sampling approach} \label{sec:Supp.Sampling}

A Gibbs step is used to sample $Y_i$ given $\mathbf Z_i$ and $\mathbf C_i$. Using Bayes rule, assuming conditional independence of $Z_i$ and $C_i$ given $Y_i$, and independence between sequences given the parameters, for $i=1, \dots, N$,
\begin{equation} \label{eq:Y_Gibbs}
\begin{aligned}
    &\Prob (Y_i=y | \mathbf Z_i, \mathbf C_i, \beta_0, \boldsymbol \beta, \boldsymbol \theta, \boldsymbol \nu, \boldsymbol \phi, \boldsymbol \alpha, \omega_0, \boldsymbol \omega) \\
    &\quad\quad = \frac{f_Y(Y_i=y | \beta_0, \boldsymbol \beta, \boldsymbol \theta) f_Z(\mathbf Z_i | Y_i=y, \boldsymbol \nu, \boldsymbol \phi) f_{\mathbf C}(\mathbf C_i | Y_i=y, \boldsymbol \alpha, \omega_0, \boldsymbol \omega) }
    {\sum_{\ell=1}^L f_Y(Y_i=\ell | \beta_0, \boldsymbol \beta, \boldsymbol \theta) f_Z(\mathbf Z_i | Y_i=\ell, \boldsymbol \nu, \boldsymbol \phi) f_{\mathbf C}(\mathbf C_i | Y_i=\ell, \boldsymbol \alpha, \omega_0, \boldsymbol \omega)}    
\end{aligned}
\end{equation}
where 
$$f_Z(\mathbf Z_i | Y_i, \boldsymbol \nu, \boldsymbol \phi) = \prod_{k \in \boldsymbol \Lambda^Y_i} f_Z(Z_{i,k} | Y_i, \boldsymbol \nu, \boldsymbol \phi)$$ 
and 
$$f_C(\mathbf C_i | Y_i, \boldsymbol \alpha, \omega_0, \boldsymbol \omega) = \prod_{k \in \boldsymbol \Lambda^C_i} f_C(\mathbf C_{i,k} | Y_i, \boldsymbol \alpha, \omega_0, \boldsymbol \omega).$$ 

Each parameter vector $\boldsymbol \beta$, $\tilde{\boldsymbol \theta}$, and $\boldsymbol \omega$ is sampled in block. Parameter vectors $\tilde{\boldsymbol \phi}_y$ and $\boldsymbol \alpha_y$ are sampled together for each category $y$, while each element of $\boldsymbol \nu$ and $\tilde{\boldsymbol \nu}$ are sampled univariately.

\subsection{Ranked probability score} \label{sec:Supp.RPS}

To compare model fits, we use the ranked probability score (RPS) as described by \cite{Gneiting2007}. It is a proper scoring rule, meaning that it rewards both calibration and sharpness. For our purposes this means that it rewards statistical consistency between the posterior and true values as well as tight credible intervals. We apply RPS using the equation
\begin{equation}
    \text{RPS}(\hat p, p) = \frac{1}{L-1} \sum_{\ell=1}^{L-1} \left( \sum_{k=1}^\ell \hat p(k) - p(k) \right)^2
\end{equation}
where $p$ and $\hat p$ are densities for a discrete random variable. Note that this corresponds to the $\ell_2$ distance between the two distributions. To summarize the RPS for each model we take the mean over all sequences.

\end{document}